\documentclass[twocolumn,letter]{jpsj2} 
\def\degangle{\kern-.2em\r{}} 

\title{
Thermal and Electrical Properties of $\gamma$-Na$_{x}$CoO$_{2}$ ($0.70\leq x\leq 0.78$)
}

\author{Hiroya \textsc{Sakurai}\thanks{E-mail: sakurai.hiroya@nims.go.jp}, Naohito \textsc{Tsujii}$^{1}$, and Eiji \textsc{Takayama-Muromachi}}

\inst{
Superconducting Materials Center, National Institute for Materials Science (SMC/NIMS), 1-1 Namiki, Tsukuba, Ibaraki 305-0044, Japan\\
$^{1}$NanoMaterials Laboratory, National Institute for Materials Science (NML/NIMS), 1-2-1 Sengen, Tsukuba, Ibaraki 305-0044, Japan
}

\abst{
We have performed specific heat and electric resistivity measurements of Na$_{x}$CoO$_{2}$ ($x=0.70$-0.78). Two anomalies have been observed in the specific heat data for $x=0.78$, corresponding to magnetic transitions at $T_{\mbox{c}}=22$ K and $T_{\mbox{k}}\simeq 9$ K reported previously. In the electrical resistivity, a steep decrease at $T_{\mbox{c}}$ and a bending-like variation at $T_{\mbox{b}}$(=120K for $x=0.78$) have been observed. Moreover, we have investigated the $x$-dependence of these parameters in detail. The physical properties of this system are very sensitive to $x$, and the inconsistent results of previous reports can be explained by a small difference in $x$. Furthermore, for a higher $x$ value, a phase separation into Na-rich and Na-poor domains occurs as we previously proposed, while for a lower $x$ value, from characteristic behaviors of the specific heat and the electrical resistivity at the low-temperature region, the system is expected to be in the vicinity of the magnetic instability which virtually exists below $x=0.70$.
}

\kword{
Na$_{x}$CoO$_{2}$, specific heat, electrical resistivity, spin fluctuation, nearly ferromagnetic
}

\begin{document}
\maketitle

Na$_{x}$CoO$_{2}$ is a quite attractive compound for condensed matter physicists and chemists because of its rich physical properties. For instance, Na$_{0.5}$CoO$_{2}$ shows unusually high thermoelectrical performance\cite{NaCo2O4N1,NaCo2O4N2}. In the case of $x=0.35$, superconductivity below $\sim$5 K is induced by water insertion\cite{NCO}. Na$_{0.75}$CoO$_{2}$ shows a magnetic transition at $T_{\mbox{c}}=22$ K, which is believed to be caused by spin density wave (SDW) formation\cite{MotohashiSDW,SugiyamaSDW1}. Moreover, for Na$_{x}$CoO$_{2}$ ($x=0.7$ - 0.75), Sommerfeld constant, $\gamma$, has been estimated from the specific heat to be $\sim$24-30 mJ/Co mol$\cdot$K$^{2}$\cite{MotohashiSDW,Ando,Bruhwiler,Miyoshi,Sales}, and the heavy fermion behavior has been observed\cite{Miyoshi}.

Despite intensive studies, there are still serious discrepancies in physical property data of Na$_{x}$CoO$_{2}$. The reports on the specific heat are inconsistent with each other, particularly those on low-temperature data: the $C/T$-$T^{2}$ curve ($C$: specific heat, $T$: temperature) in the low-temperature region has been reported to show an upturn\cite{Ando,Bruhwiler,Carretta}, linear dependence\cite{Bayrakci,Miyoshi}, or a downturn\cite{Sales}.

As regards the electrical resistivity, the variation of the data is less than that of the specific heat. When the transition at $T_{\mbox{c}}$ is magnetically observed, the resistivity shows a drop at $T_{\mbox{c}}$, while no anomaly at $T_{\mbox{c}}$ in a transition-free sample is observed. A slight bend in the resistivity around 100 K is seen clearly in the data reported by Shi $et$ $al.$\cite{Shi} and is observed in most cases\cite{Bruhwiler,Sales,Foo}, the origin of which is left unclarified.

The discrepancies in the physical properties including magnetic properties seem to be mainly caused by their strong dependencies on the Na content $x$. Very recently, we prepared $\gamma$-Na$_{x}$CoO$_{2}$ samples by varying $x$ minutely and measured their magnetic properties. We found that the solid-solution range of the system is quite narrow with $0.70\leq x\leq 0.78$, and moreover, its magnetic properties are very sensitive to $x$\cite{Sakurai}. Most previous studies were carried out only for one of the two end members with $x\sim 0.7$ and $\sim 0.78$. In the present study, we measured the specific heats and the electrical resistivities for the same sets of samples used in the previous magnetic measurements, in order to elucidate their dependences on $x$ and to clarify the origin of the above-mentioned discrepancies.

The powder samples of Na$_{x}$CoO$_{2}$ ($x=0.70$, 0.72, 0.74, 0.76, 0.78, 0.80, and 0.82) were synthesized by the conventional solid state reaction from the stoichiometric mixtures of Na$_{2}$CO$_{3}$ (99.99\%) and Co$_{3}$O$_{4}$ (99.9\%). The detailed preparation method and the results of chemical analyses of the samples are described elsewhere\cite{Sakurai}. The specific heat data were collected using a commercial physical property measurement system (PPMS, Quantum Design) for the sintered samples with weights of 14-27 $mg$. The measurements were usually carried out with decreasing temperature, and the magnetic field ($H$) was applied before cooling. The electrical resistivity ($\rho$) was also measured using PPMS by the conventional four-probe method for the sintered samples with typical dimensions of $8\times 3\times 1$ $mm^{3}$ (1-2 $mm$ distance between adjacent probes) with increasing temperature at a rate of 1 K/min, after the samples were cooled down to 1.8 K at the same rate.

\begin{figure}
\begin{center}
\includegraphics[width=7cm,keepaspectratio]{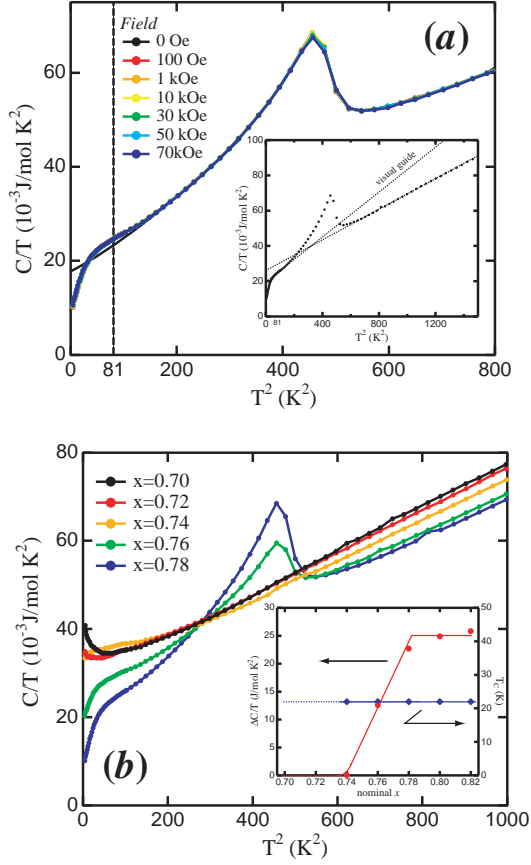}
\end{center}
\caption{
(a) $C/T$-$T^{2}$ curves at various magnetic fields. The broken and solid lines are visual guides. The inset shows $C/T$ under 0 Oe. The dotted lines are $C/T = 26.0 + 4.34 \times 10^{-2}T^{2}$ and a visual guide. (b) $C/T$-$T^{2}$ curves of Na$_{x}$CoO$_{2}$ with various $x$ values. The inset shows the $x$-dependences of $\Delta C/T$ (red markers) and $T_{\mbox{c}}$ (blue markers). The solid lines are visual guides.
}
\label{Cp}
\end{figure}

The $C/T$-$T^{2}$ curves of Na$_{0.78}$CoO$_{2}$ are shown in Fig. \ref{Cp}(a). Two anomalies are seen: (i) a sharp peak at $T_{\mbox{c}}=22$ K and (ii) a dull downward bend at $T_{\mbox{k}}\simeq 9$ K. These anomalies correspond to the magnetic transitions\cite{Sakurai}. The latter transition will be discussed later. The $C/T$-$T^{2}$ curves of Na$_{x}$CoO$_{2}$ with various $x$ values are shown in Fig. \ref{Cp}(b).   

The peak position and shape at $T_{\mbox{c}}$ measured with increasing temperature were completely the same as those measured with decreasing temperature, and, as seen in Fig. \ref{Cp}(a), $T_{\mbox{c}}$ is also independent of $H$. Moreover, it does not depend on $x$ as shown in Fig. \ref{Cp}(b). Only the jump in $C/T$ at $T_{\mbox{c}}$, $\Delta C/T$, decreases with decreasing $x$ below 0.78 as seen in the inset of Fig. \ref{Cp}(b). These strongly suggest that the phase responsible for the transition in question does not change but only its fraction varies with $x$, which is consistent with the magnetic measurements\cite{Sakurai}. Namely, for the samples with $x$ above 0.74, a phase separation into Na-rich and Na-poor domains occurs. In addition, as seen in the inset of Fig. \ref{Cp}(b), the independent behavior of $\Delta C/T$ on $x$ for $x>0.78$ supports the higher limit of the solid-solution range of $x=0.78$\cite{Sakurai}.

\begin{figure}
\begin{center}
\includegraphics[width=7cm,keepaspectratio]{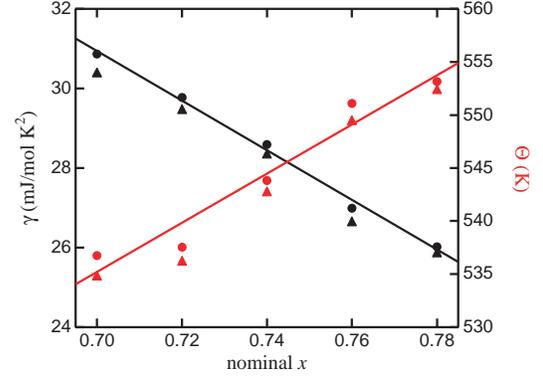}
\end{center}
\caption{
$x$-dependences of $\gamma$ (black markers) and $\Theta$ (red markers). The circular and triangular markers represent the values estimated using eqs. \ref{LinearC} and \ref{Debye}, respectively. The black and red lines are $\gamma =74.6-62.4x$ and $\Theta =373+232x$, respectively. }
\label{Para}
\end{figure}

The Sommerfeld constants $\gamma$ and Debye temperatures $\Theta$ were first estimated by the function of
\begin{equation}
C/T=\gamma+AT^{2}
\label{LinearC}
\end{equation}
($A=\frac{12\pi^{4}}{5}\frac{Nk_{B}}{\Theta^{3}}$: $N$, the number of the atom and $k_{B}$, Boltzmann constant) using the data between 26 and 36 K ($700\leq T^{2} \leq 1300$ K$^{2}$). The parameters obtained are shown in Fig. \ref{Para} by circular markers. Since the temperature range seems to be high, the same parameters were estimated from the same data by a different function:
\begin{equation}
C=\gamma T + 9Nk_{B}(\frac{T}{\Theta})^{3}\int^{\Theta/T}_{0}dx\frac{x^{4}e^{x}}{(e^{x}-1)^{2}} \mbox{ (Debye model), }
\label{Debye}
\end{equation}
where the fitting parameters are only $\gamma$ and $\Theta$. As shown in Fig. \ref{Para}, the $\gamma$ and $\Theta$ values obtained from the two different equations are in good agreement with each other, which means that eq. \ref{LinearC} is applicable to this temperature range of these compounds. These $\gamma$ values agree well with those reported previously\cite{MotohashiSDW,Ando,Bruhwiler,Miyoshi,Sales}, and lie on the line of $\gamma = 74.6-62.4x$. This $x$-dependence of $\gamma$ indicates that the density of states (DOS) decreases with increasing Fermi energy, although it is difficult to estimate the influence of the phase separation of the samples with higher $x$ on the $\gamma$ values; it is confirmed that this negative inclination corresponds to that of the energy dependence of DOS above the Fermi energy of Na$_{0.5}$CoO$_{2}$\cite{Singh}.

Below about 15 K, the line of eq. \ref{LinearC} is located above the experimental data (see the inset of Fig. \ref{Cp}(a)), which implies that some part of DOS is lost due to the transition at $T_{\mbox{c}}$. This fact seems to be in favor of the SDW formation at $T_{\mbox{c}}$ rather than the typical second-order transition suggested previously\cite{MotohashiSDW}. However, it should be noted that the SDW state is realized only in a part of a sample as mentioned above and suggested previously\cite{Sakurai}.

The dull anomaly at $T_{\mbox{k}}$, which has been observed by a single crystal\cite{Sales}, is likely due to another weak ferromagnetic transition\cite{Sakurai}. The transition temperature $T_{\mbox{k}}$ is independent of $H$ as in the case of $T_{\mbox{c}}$. Since the degree of the anomaly seems to change synchronically with $\Delta C/T$ at $T_{\mbox{c}}$, it seems that the domain which undergoes the transition at $T_{\mbox{c}}$ is followed by the transition at $T_{\mbox{k}}$. The transition at $T_{\mbox{k}}$ may be caused by the change in the magnetic structure formed in the first transition at $T_{\mbox{c}}$ with additional lost of DOS. It is, however, necessary to perform microscopic experiments, such as nuclear magnetic resonance (NMR), to elucidate details of the transition.

The $T$-dependence of $\rho$ for $x=0.78$ is shown in Fig. \ref{rho}(a). Three characteristic features are seen in this log-log plot: (i) a metallic behavior in the entire $T$-range, (ii) a steep decrease in $\rho$ below $T_{\mbox{c}}$, and (iii) bending of the curve at $T_{\mbox{b}}=120$ K. The metallic behavior even below $T_{\mbox{c}}$ and $T_{\mbox{k}}$ is consistent with the existence of the residual $\gamma$ of approximately 10 mJ/Co mol$\cdot$K$^{2}$ for $x=0.78$ at 0 K.

\begin{figure}
\caption{
(a) Temperature dependences of $\rho$ for $x=0.78$ shown on the logarithmic scale. The black and red lines represent the raw data and the data compensated for the residual resistivity, respectively. The broken lines indicate $T^{\alpha}$-dependences with $\alpha_{\mbox{mid}}=0.372$ and $\alpha_{\mbox{high}}=0. 833$. The dotted lines indicate $T^{\alpha}$-dependences with $\alpha_{\mbox{low}}=1.46$, $\alpha_{\mbox{mid}}=0.558$, and $\alpha_{\mbox{high}}=0.992$. (b) Temperature dependences of $\rho(T)/\rho(1.8K)$ of Na$_{x}$CoO$_{2}$ with various $x$ values shown on the logarithmic scale.
}
\label{rho}
\end{figure}

The steep decrease in $\rho$ below $T_{\mbox{c}}$ is, of course, related to the magnetic transition seen in the specific heat and the magnetic susceptibility\cite{Sakurai}. Indeed, as shown in Fig. \ref{rho}(b), this anomaly becomes less pronounced with decreasing $x$ in consistent with the specific heat and magnetic susceptibility data. Therefore, this anomaly reflects the intrinsic nature of the domain which undergoes the magnetic transition at $T_{\mbox{c}}$.  In the case of a lower $x$ value, a marked decrease in $\rho$ below 40 K, which is seen even for $x=0.78$, seems to continue down to 1.8 K as reported by Miyoshi $et$ $al$.\cite{Miyoshi}, the origin of which is unknown.

The bending-like variation of $\rho$ at $T_{\mbox{b}}$ has been seen in the in-plane resistivity measured by a single crystal\cite{NaCo2O4N1,Bruhwiler,Foo}, which means that the bend is intrinsic. To explain this behavior, the data above 30 K were fitted by the equation $\rho=AT^{2}\ln (E_{F}/T)$ ($E_{F}$: Fermi energy), which is based on a two-dimensional (2D) Fermi gas model\cite{Hodges,Bloom}. The resulting function seemed to reproduce the data to some extent, but $E_{F}$ was very small, being about 700 K. Then, we calculated the exponents $\alpha$ in $\rho\propto T^{\alpha}$ for the two temperature ranges 40-70 K  ($\alpha _{\mbox{mid}}$) and 150-210 K ($\alpha _{\mbox{high}}$) for $x=0.78$ to obtain $\alpha _{\mbox{mid}}=0.371$ and $\alpha _{\mbox{high}}=0.833$. To eliminate the influence of residual resistivity, the residual resistivity, which was estimated to be 1.36 m$\Omega\cdot cm$ from the data below 10 K by a trinomial, was subtracted from the raw data. For the data compensated for the residual resistivity, the exponents become $\alpha _{\mbox{mid}}=0.558$ and $\alpha _{\mbox{high}}=0.992$ (see Fig. \ref{rho}(a)). Thus, $\rho$ obeys the almost $T$-linear relation in the high temperature range as the high-$T_{\mbox{c}}$ cuprates do. The optical spectra for $x=0.7$ also suggest a phenomenon similar to that of the high-$T_{\mbox{c}}$ cuprates\cite{Hwang}. On the other hand, the compensated data gave the $T^{1.46}$-dependence for the temperature range below $T_{\mbox{c}}$. From these facts, a typical Fermi liquid behavior was not seen at least for $x=0.78$ in any $T$-range and by any procedure. In cases of $x=0.74$ and 0.76, the $T^{2}$-dependence was seen but only below 5 K, which is consistent with the data obtained by a single crystal\cite{Miyoshi}.

\begin{figure}
\begin{center}
\includegraphics[width=7cm,keepaspectratio]{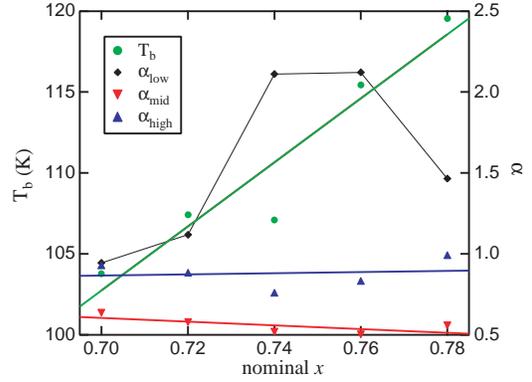}
\end{center}
\caption{
$x$-dependences of $T_{\mbox{b}}$, $\alpha _{\mbox{low}}$, $\alpha _{\mbox{mid}}$ and $\alpha _{\mbox{high}}$ estimated from the data compensated for the residual resistivities from the raw data. The solid lines represent $T_{\mbox{b}}=-35.8+198x$, $\alpha _{\mbox{mid}}=1.41-1.15x$, and $\alpha _{\mbox{high}}=0.610-0.365x$, respectively.
}
\label{rhoPara}
\end{figure}

The $\alpha _{\mbox{low}}$, $\alpha _{\mbox{mid}}$ and $\alpha _{\mbox{high}}$ values were determined for every $x$ using the data compensated for the residual resistivities as shown in Fig. \ref{rhoPara}. $T_{\mbox{b}}$ was determined from the crossing point of the two lines with $\alpha _{\mbox{mid}}$ and $\alpha _{\mbox{high}}$ and is plotted in Fig. \ref{rhoPara}. $\alpha _{\mbox{high}}$ is almost independent of $x$ and is close to unity, indicating the $T$-linear-like behavior in the high $T$-range. On the other hand, $T_{\mbox{b}}$ increases linearly with $x$. Similar tendencies in $\alpha _{\mbox{mid}}$, $\alpha _{\mbox{high}}$, and $T_{\mbox{b}}$ were obtained when we used the raw data. Since no anomaly is seen around $T_{\mbox{b}}$ in the specific heat and the magnetic susceptibility\cite{Sakurai}, the bend is not due to a transition. However, at the present stage, the physical meaning of $T_{\mbox{b}}$ is unclear, although the anomalies at approximately $T_{\mbox{b}}$ have been observed by angle-resolved photoemission spectroscopy and optical measurements\cite{Hasan,Wang}. Similar bends have been observed in the resistivity of LiV$_{2}$O$_{4}$ ($T_{\mbox{b}}\sim 300$-400 K\cite{LiV2O4}) and in-plane resistivities of Sr$_{n+1}$Ru$_{n}$O$_{3n+1}$ ($T_{\mbox{b}}\sim 250$ K for $n=1$\cite{Sr2RuO4} and $T_{\mbox{b}}\sim 200$ K for $n=2$\cite{Sr3Ru2O7}). Thus, since these compounds have been expected to have a strong orbital or spin fluctuation, the bend of Na$_{x}$CoO$_{2}$ may be due to some kind of fluctuation, such as an orbital or spin fluctuation.

Finally, we will discuss about the physical properties for $x=0.70$. The resistivity of this compound shows a $T^{\alpha}$-dependence with $\alpha _{\mbox{low}}\sim 1$ below 7 K, and the $\alpha _{\mbox{low}}$ value increased up to 1.4 with increasing magnetic field up to 7 T, which is consistent with the results measured by a single crystal\cite{Li}. Since no magnetic transition is observed in the specific heat and magnetic measurements\cite{Sakurai}, this behavior of the electrical resistivity suggests that this compound is in the vicinity of the magnetic instability which virtually exists below $x=0.70$. This idea seems to be consistent with the enhancement of $C/T$ below 10 K, because $\gamma$ of a nearly ferromagnetic or nearly antiferromagnetic compound is often enhanced by the spin fluctuation\cite{Moriya}. Since the Wilson ratio $R=\frac{\pi^{2}}{3}(\frac{k_{B}}{\mu _{B}})^{2}\frac{\chi}{\gamma}$ ($\mu _{B}$: Bohr magneton and $\chi$: magnetic susceptibility) is calculated to be $R=2.8$-2.9 for $x=0.70$, using $\chi =1.25$-$1.17\times 10^{-3}$ emu/mol at 26-36 K, the spin fluctuation seems to be ferromagnetic. The band calculation also indicates this ferromagnetic tendency\cite{Singh}.

In summary, we have performed the specific heat and resistivity measurements of Na$_{x}$CoO$_{2}$ ($x=0.70$-0.78) for the same sets of samples used in the previous magnetic measurements\cite{Sakurai}. In the specific heat for $x=0.78$, two anomalies were seen, corresponding to magnetic transitions at $T_{\mbox{c}}=22$ K and $T_{\mbox{k}}\simeq 9$ K. Both anomalies become less pronounced simultaneously with decreasing $x$, while keeping $T_{\mbox{c}}$ and $T_{\mbox{k}}$ unchanged, and disappear below $x=0.72$. This behavior is consistent with our phase separation model proposed previously. The Sommerfeld constant was estimated to be $\gamma=26$-31 mJ/mol$\cdot$K$^{2}$, which is consistent with the previous reports. The resistivity measurements showed that this system is metallic for both the entire $x$ and $T$ ranges. The steep decrease at $T_{\mbox{c}}$ and the bending-like variation at $T_{\mbox{b}}$ (=120K) were found in the resistivity for $x=0.78$. $T_{\mbox{b}}$ increased slightly with $x$, the origin of which is unclear. From these results, for a higher $x$ value, a phase separation into Na-rich and Na-poor domains occurs as we previously proposed, while, for a lower $x$ value, the system is expected to be in the vicinity of the magnetic instability which virtually exists below $x=0.70$.
 
\section*{Acknowledge}
Special thanks to S. Takenouchi (NIMS) for chemical analysis. We would like to thank K. Takada, T. Sasaki, A. Tanaka, M. Kohno (NIMS), and K. Ishida (Kyoto University) for fruitful discussion. This study was partially supported by a Grants-in-Aid for Scientific Research (B) from Japan Society for the Promotion of Science (16340111). One of the authors (H.S) is a Research Fellow of the Japan Society for the Promotion of Science.

\end{document}